\documentclass[10pt,journal,cspaper,compsoc]{IEEEtran}
\usepackage{graphicx}
\usepackage{float}
\usepackage{multirow}
\usepackage{amsmath}
\usepackage{cite}
\usepackage{subfigure}
\usepackage{url}
\usepackage{hyperref}

\usepackage{array}
\newcommand{\PreserveBackslash}[1]{\let\temp=\\#1\let\\=\temp}
\newcolumntype{C}[1]{>{\PreserveBackslash\centering}p{#1}}
\newcolumntype{R}[1]{>{\PreserveBackslash\raggedleft}p{#1}}
\newcolumntype{L}[1]{>{\PreserveBackslash\raggedright}p{#1}}

\ifCLASSINFOpdf
\else
\fi
\begin{document}
%
\title{Identifying Stable Patterns over Time for Emotion Recognition from EEG}
%
%
%
%

\author{Wei-Long~Zheng,
        Jia-Yi~Zhu,
        and~Bao-Liang~Lu*,~\IEEEmembership{Senior~Member,~IEEE}
\IEEEcompsocitemizethanks{\IEEEcompsocthanksitem Wei-Long Zheng, Jia-Yi Zhu and Bao-Liang Lu are with the Center for Brain-Like Computing and Machine Intelligence, Department of Computer Science and Engineering, Shanghai Jiao Tong University and Key Laboratory of Shanghai Education Commission for Intelligent Interaction and Cognitive Engineering, Shanghai Jiao Tong University, 800 Dong Chuan Road, Shanghai 200240, China.\protect\\
E-mail: \{weilonglive,zhujiayi1991\}@gmail.com, bllu@sjtu.edu.cn.\hfil\break *Corresponding author}
\thanks{}}

\IEEEcompsoctitleabstractindextext{%
\begin{abstract}
In this paper, we investigate stable patterns of electroencephalogram (EEG) over time for emotion recognition using a machine learning approach. Up to now, various findings of activated patterns associated with different emotions have been reported. However, their stability over time has not been fully investigated yet. In this paper, we focus on identifying EEG stability in emotion recognition. To validate the efficiency of the machine learning algorithms used in this study, we systematically evaluate the performance of various popular feature extraction, feature selection, feature smoothing and pattern classification methods with the DEAP dataset and a newly developed dataset for this study. The experimental results indicate that stable patterns exhibit consistency across sessions; the lateral temporal areas activate more for positive emotion than negative one in beta and gamma bands; the neural patterns of neutral emotion have higher alpha responses at parietal and occipital sites; and for negative emotion, the neural patterns have significant higher delta responses at parietal and occipital sites and higher gamma responses at prefrontal sites. The performance of our emotion recognition system shows that the neural patterns are relatively stable within and between sessions.

\end{abstract}

}

\maketitle

\IEEEdisplaynotcompsoctitleabstractindextext

%
\IEEEpeerreviewmaketitle

\section{Introduction}

\IEEEPARstart{E}{MOTION} plays an important role in human communication and decision making. Although it seems natural to us with emotions in our daily life, we have little knowledge on the mechanisms of emotional function of the brain and modeling human emotion. In recent years, the research on emotion recognition from EEG has attracted great interest from a vast amount of interdisciplinary fields from psychology to engineering, including basic studies on emotion theories and applications to affective Brain-Computer Interaction (aBCI) \cite{muhl2014survey}, which enhances the BCI systems with the ability to detect, process, and respond to users affective states using physiological signals.

Although many progresses in theories, methods and experiments that support affective computing have been made in the past several years, the problem of detecting and modeling human emotions in aBCI remains largely unexplored \cite{muhl2014survey}. Emotion recognition is critical because computers can never respond to users' emotional states without recognizing human emotions. However, emotion recognition from EEG is very challenging due to the fuzzy boundaries and individual different variations of emotion. In addition, we can't obtain the `ground truth' of human emotions in theory, that is, the true label of EEG corresponding different emotional states. The reason is that emotion is considered as a function of time, context, space, language, culture, and races \cite{kim2008emotion}.

Many previous studies focus on subject-dependent and subject-independent patterns and evaluations for emotion recognition. However, the stable patterns and performance of models over time are not fully exploited, which are very important for real world applications. Stable EEG patterns are considered as neural activities such as critical brain areas and critical frequency bands that share commonality across individuals and sessions under different emotional states. Although task-related EEG is sensitive to change due to different cognitive states and environmental variables \cite{mcevoy2000test}, we intuitively think that the stable patterns for specific tasks should exhibit consistency among repeated sessions of the same subjects. In this paper, we focus on the following issues of EEG-based emotion recognition: What is the capability of EEG signals for discrimination of different emotions? Are there any stable EEG patterns of neural oscillations or brain regions for representing emotions? What is the performance of the models based on machine learning approach from day to day?

The main contributions of this paper to emotion recognition from EEG can be summarized as follows: 1) We have developed a novel emotion EEG dataset as a subset of SEED (SJTU Emotion EEG Dataset), that will be publicly available for research to evaluate stable patterns across subjects and sessions. To the best of our knowledge, there is no available public EEG dataset for the analysis of stability of neural patterns regarding emotion recognition. 2) We carry out a systematic comparison and a qualitative evaluation of different feature extraction, feature selection, feature smoothing and pattern classification methods on a public available EEG dataset, DEAP, and our own dataset, SEED. 3) We adopt discriminative Graph regularized Extreme Learning Machine (GELM) to identify the stable patterns over time and evaluate the stability of our emotion recognition model with cross-session schemes. 4) Our experiment results reveal that neural signatures for three emotions (positive, neutral and negative) do exist and the EEG patterns at critical frequency bands and brain regions are relatively stable within and between sessions.

The layout of the paper is as follows. In Section 2, we give a brief overview of related work on emotion recognition from EEG, as well as the findings of stable patterns for different emotions. A systematic description of brain signal analysis methods and classification procedure for feature extraction, dimensionality reduction and classifiers is given in Section 3. Section 4 presents the motivation and rationale for our emotion experimental setting. A detailed explanation of all the materials and protocol we used is described. A systematic evaluation on different methods is conducted on the DEAP dataset and our SEED dataset. We use time-frequency analysis to find the neural signatures and stable patterns for different emotions, and evaluate the stability of our emotion recognition models over time. In Section 5, we make a conclusion about our work.

\begin{table*}
\caption{Various studies on emotion classification using EEG and the best performance reported in each study}\label{tab:table1}
\renewcommand{\arraystretch}{1.3}
\begin{center}
\begin{tabular}{|c|p{3em}|p{3em}|L{5.cm}|L{2.5cm}|L{2.4cm}|p{1cm}|}
\hline
\hline
Study     & Stimuli    & \#Chan. & Method Description                                                                            & Emotion states                             & Accuracy              & Pattern Study                                      \\ \hline
\cite{bos2006eeg}  & IAPS,\newline IADS & 3         & Power of alpha and beta, then PCA, 5 subjects, classification with FDA                        & Valence and arousal                        & Valence: 92.3\%, arousal: 92.3\%            &$\times$                                       \\ \hline
\cite{heraz2007predicting}  & IAPS       & 2         & Amplitudes of four frequency bands, 17 subjects, evaluated KNN, Bagging                       & Valence (12), arousal (12) and dominance (12)             & Valence: 74\%, arousal: 74\%, and dominance: 75\%           &$\times$\\ \hline
\cite{murugappan2010classification}  & Video      & 62        & Wavelet features of alpha, beta and gamma, 20 subjects, classification with KNN and LDA       & disgust, happy,\newline surprise, fear and neutral & 83.26\%          &$\times$                                          \\ \hline
\cite{lin2010eeg}  & Music      & 24        & Power spectral density and asymmetry features of five frequency bands, 26 subjects, evaluated SVM                & Joy, anger, sadness, and pleasure          & 82.29\%                          &$\surd$                          \\ \hline
\cite{brown2011towards}  & IAPS       & 8         & Spectral power features, 11 subjects, KNN                                                     & Positive, negative and neutral             & 85\%                                  &$\times$                     \\ \hline
\cite{petrantonakis2011novel}  & IAPS       & 4         & Asymmetry index of alpha and beta power, 16 subjects, SVM                                     & Four quadrants of the valence-arousal space                           & 94.4\% (subject-dependent), 62.58\% (subject-independent)                     &$\times$                                \\ \hline
\cite{koelstra2012deap}  & Video      & 32        & Spectral power features of five frequency bands, 32 subjects, Gaussian naive Bayes classifier & Valence (2), arousal (2) and liking (2)          & Valence: 57.6\%, arousal: 62\% and liking: 55.4\%  &$\times$\\ \hline
\cite{hadjidimitriou2012toward}  & Music      & 14        & Time-frequency (TF) analysis, 9 subjects, KNN, QDA and SVM                                    & Like and dislike                           & 86.52\%                    &$\surd$                                \\ \hline
\cite{soleymani2012multimodal}  & Video      & 32        & Power spectral density features of five frequency bands, modality fusion with eye track, 24 subjects, SVM        & Valence (3) and arousal (3)                    & Valence: 68.5\%, arousal: 76.4\%     &$\times$             \\ \hline
\cite{wang2013emotional}  & Video      & 62        & Power spectrum features, wavelet features, nonlinear dynamical features, 6 subjects, SVM      & Positive and negative                      & 87.53\%                              &$\surd$                      \\ \hline
\cite{jenke2014feature}  & IAPS      & 64        & Higher Order Crossings, Higher Order Spectra and Hilbert-Huang Spectrum features, 16 subjects, QDA        & Happy, curious, angry, sad, quiet                    & 36.8\%     &$\surd$             \\ \hline
\hline
\end{tabular}
\begin{quote}
IAPS and IADS denotes the International Affective Picture System and the International Affective Digital Sounds, respectively. The numbers given in parenthesis denote the numbers of categories for each dimension. Pattern study denotes revealing neural activities (critical brain areas and critical frequency bands) that share commonality across subjects or sessions. Classifiers include $K$ Nearest Neighbors (KNN), Fisher's Discriminant Analysis (FDA), Linear Discriminant Analysis (LDA), Quadratic Discriminant Analysis (QDA), Bagging, and Support Vector Machine (SVM).
\end{quote}
\end{center}
\end{table*}

\section{Related Work}
\subsection{Emotion Recognition Methods}
In the field of affective computing, a vast amount of studies has been conducted toward emotion recognition based on different signals. Many efforts have been made to recognize emotions using external emotion expression, such as facial expression \cite{zeng2009survey}, speech \cite{schuller2005meta}, and gestures \cite{d2009automatic}. However, sometimes the emotional states remain internal and cannot be detected by external expression. Considering extreme cases where people do not say anything but actually they are angry, even smile during negative emotional states due to the social masking \cite{ekman1989argument}. In these cases, the external emotional expression can be controlled subjectively and these external cues for emotion recognition may be inadequate.

Emotion is an experience that is associated with a particular pattern of physiological activity including central nervous system and autonomic nervous system. Contrary to audiovisual expression, it is more objective and straightforward to recognize emotions using physiological activities. These physiological activities can be recorded by noninvasive sensors, mostly as electrical signals. These measures include skin conductivity, electrocardiogram, electromyogram, electrooculogram, and EEG. A detailed review of emotion recognition methods can be found in \cite{calvo2010affect}.

With the fast development of micro-nano technologies and embedded systems, it is now conceivable to port aBCI systems from laboratory to real-world environments. Many advanced dry electrodes and embedded systems are developed to handle the wearability, portability, and practical use of these systems in real world applications \cite{lin2011novel, grozea2011bristle}. Various studies in affective computing community try to build computational models to estimate emotional states based on EEG features. In short, a brief summary of emotion recognition using EEG is presented in Table~\ref{tab:table1}. These studies show the efficiency and feasibility of building computational models of emotion recognition using EEG. In these studies, the stimuli used in emotion recognition experiments contains still images, music and videos and emotions evaluated in most studies are discrete.

\subsection{EEG Patterns Associated with Emotions}
One of the goals in affective neuroscience is to examine whether patterns of brain activity for specific emotions exist, and whether these patterns are to some extent common across individuals. Various studies have examined the neural correlations of emotions. It seems that there don't exist any processing modules for specific emotion. However, there may exist neural signatures of specific emotion, as a distributed pattern of brain activity \cite{kassam2013identifying}. Mauss and Robinson \cite{mauss2009measures} proposed that emotional state is likely to involve circuits rather than any brain region considered in isolation. To AC researchers, to identify neural patterns that are both common across subjects and stable across sessions can provide valuable information for emotion recognition from EEG.

Cortical activity in response to emotional cues was related to the lateralization effect. Muller \emph{et al.} \cite{muller1999processing} reported increased gamma (30-50Hz) power for negative valence over left temporal region. Davidson \emph{et al.} {\cite{davidson1982asymmetrical,davidson1992anterior}} showed that frontal EEG asymmetry is hypothesized to relate to approach and withdrawal emotions, with heightened approach tendencies reflected in left frontal activity and heightened withdrawal tendencies reflected in relative right-frontal activity. Nie \emph{et al.} \cite{nie2011eeg} reported that the subject-independent features associated with positive and negative emotions are mainly in the right occipital lobe and parietal lobe for the alpha band, the central site for beta band, and the left frontal lobe and right temporal lobe for gamma band. Balconi \emph{et al.} \cite{balconi2009appetitive} found that frequency band modulations are effected by valence and arousal rating, with an increased response for high arousing and negative or positive stimuli in comparison with low arousing and neutral stimuli.

For EEG-based emotion recognition, subject-dependent and subject-independent schemes are always used for evaluating the performance of emotion recognition systems. As shown in Table~\ref{tab:table1}, some findings of activated patterns such as critical channels and oscillations associated with different emotions have been proposed. However, a major limitation is that they extract activated patterns only across subject but do not consider the time factor.

The studies of internal consistency and test-retest stability of EEG can be dated back to many years ago \cite{salinsky1991test,mcevoy2000test,gudmundsson2007reliability}, especially for clinical applications \cite{allen2004stability}. McEvoy \emph{et al.} \cite{mcevoy2000test} proposed that under appropriate conditions, task-related EEG has sufficient retest reliability for use in assessing clinical changes. However, these previous studies investigated the stability of EEG features under different conditions, for example, a working memory task \cite{mcevoy2000test}. Moreover, in these studies, stability and reliability are often quantified using statistical parameters such as intraclass correlation coefficients \cite{gudmundsson2007reliability}, instead of the performance of pattern classifiers.

So far, a few preliminary studies on stability and reliability of neural patterns for emotion recognition have been conducted. Lan \emph{et al.} \cite{lan2014stability} presented a pilot study of stability of features in emotion recognition algorithms. However, in their stability assessment, the same features derived from the same channel from the same emotion class of the same subject were grouped together to compute the correlation coefficients. Furthermore, their experiments were conducted on a small group of subjects with 14-channel EEG signals. They investigated the stability of each feature instead of neural patterns that we focus on in this paper. Up to now, there is no systematic evaluation about the stability for activated patterns over time in previous studies. The performance of emotion recognition systems over time is still an unsolved problem for developing real-world application systems. Therefore, our major aim in this paper is to investigate the stable EEG patterns over time using time frequency analysis and machine learning approaches. Here, we want to emphasize that we do not study neural patterns under emotion regulation \cite{gross2009handbook}, but for specific emotional states during different times.

To investigate various critical problems of emotion recognition from EEG, we face a serious lack of publicly available emotional EEG datasets. To the best of our knowledge, the only publicly available emotional EEG datasets are MAHNOB HCI \cite{soleymani2012multimodal} and DEAP \cite{koelstra2012deap}. The first one includes EEG, physiological signals, eye gaze, audio, and facial expressions of 30 people when watching 20 emotional videos. The DEAP dataset includes the EEG and peripheral physiological signals of 32 participants when watching 40 one-minute music videos. It also contains participants' rate of each video in terms of the levels of arousal, valence, like/dislike, dominance, and familiarity. However, these datasets do not contain EEG data from different sessions for the same subject, which can not be used for investigating the stable patterns over time. Since there is no available published EEG dataset for the analysis of stability of neural patterns for emotion recognition, we develop a new emotional EEG dataset for this study as a subset of SEED\footnote{\url{http://bcmi.sjtu.edu.cn/~seed/}}.

\section{Emotion Experiment Design}
In order to investigate neural signatures of different emotions and stable patterns over time, we design a new emotion experiment to collect EEG data, which is different from other existing publicly available datasets. In our experiments, the same subject performs the emotion experiments three times with an interval of one week or longer. We choose film clips as emotion elicitation materials. These emotional films contain both scene and audio, which can expose subjects to more real-life scenarios and elicit strong subjective and physiological changes.

In our emotion experiments, Chinese film clips are used considering that native culture factors may affect elicitation in emotion experiments \cite{zhengmultimodal}. In the preliminary study, we selected a pool of emotional film clips from famous Chinese films manually. Twenty participants were asked to assess their emotions when watching the selected film clips by scores (1-5) and keywords (positive, neutral and negative). The criteria for selecting film clips are as follows: (a) the length of the whole experiment should not be too long in case it will make subjects visual fatigue; (b) the videos should be understood without explanation; and (c) the videos should elicit a single desired target emotion. At last, 15 Chinese film clips for positive, neutral and negative emotions were chosen from the pool of materials, which received 3 or higher sores of mean ratings from the participants. Each emotion has five film clips in one experiment. The duration of each film clip is about 4 minutes. Each film clip is well edited to create coherent emotion eliciting. The details of the film clips used in the experiments are listed in Table~\ref{tab:table2}.

\begin{table}[h]\normalsize
\caption{Details of film clips used in our emotion experiment}\label{tab:table2}
\begin{center}
\begin{tabular}{p{1em}p{3.1em}p{12.5em}p{1.8em}}
\hline
\hline
No. & Labels & Film clips sources   & \#clips      \\ \hline
1   & negative      & Tangshan Earthquake        &2\\ \hline
2   & negative      & Back to 1942               &3        \\ \hline
3   & positive      & Lost in Thailand           &2\\ \hline
4   & positive      & Flirting Scholar           &1\\ \hline
5   & positive      & Just Another Pandora's Box &2\\ \hline
6   & neutral       & World Heritage in China    &5\\ \hline
\hline
\end{tabular}
\end{center}
\end{table}

Fifteen subjects (7 males and 8 females; mean: 23.27, std: 2.37) who are different from those in the preliminary study, participate in the experiments. In order to investigate neural signatures and stable patterns across sessions and individuals, each subject is required to perform the experiments for three sessions. The time interval between two sessions is one week or longer. All participants are native Chinese students from Shanghai Jiao Tong University with self-reported normal or corrected-to-normal vision and normal hearing. Before the experiments, the participants are informed about the experiment and instructed to sit comfortably, watch the forthcoming movie clips attentively without diverting their attention from the screen, and refrain as much as possible from overt movements.

\begin{figure}[!t]
\centering
\centerline{\includegraphics[width=2.in]{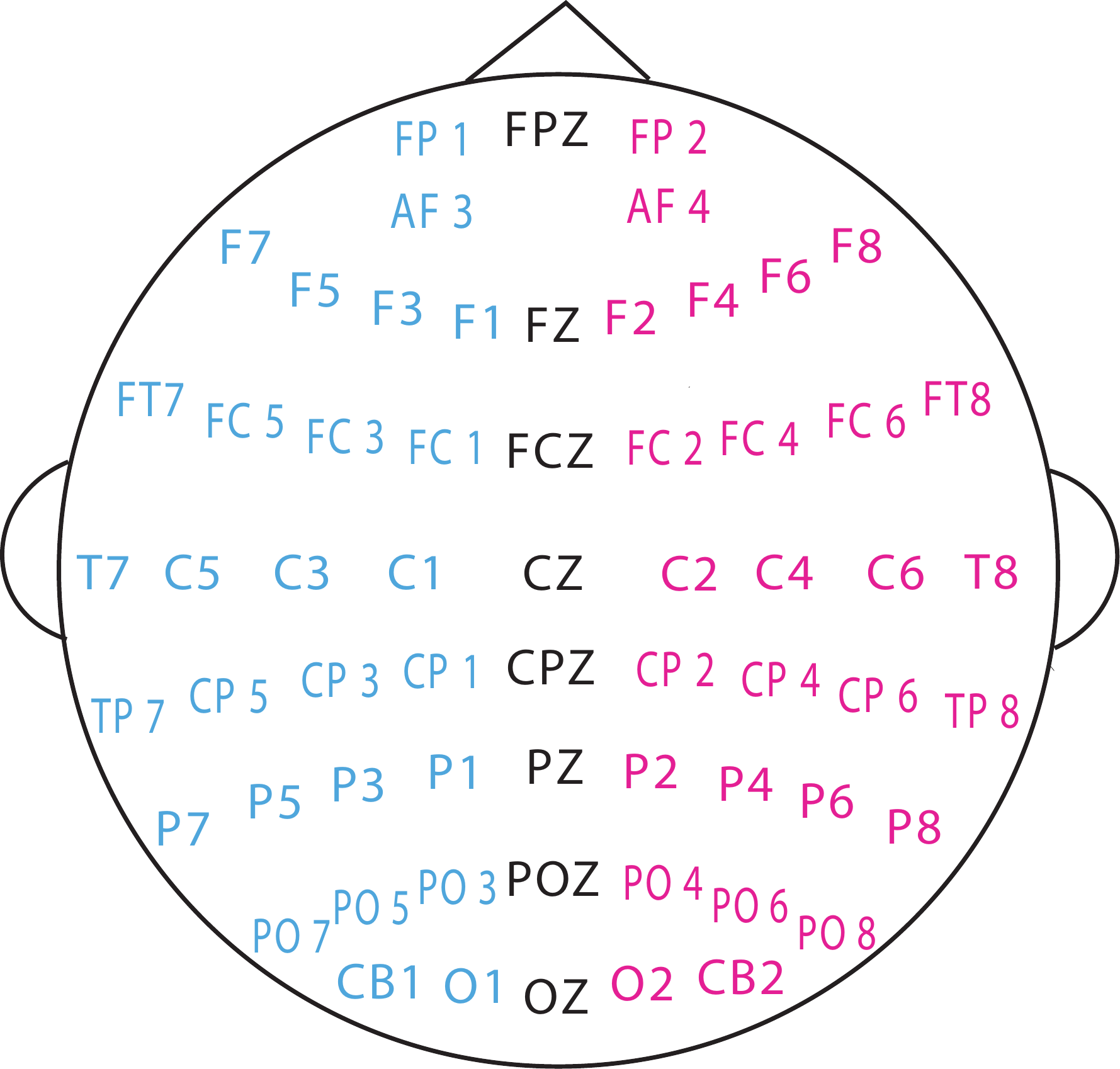}}
\caption{The EEG cap layout for 62 channels}\label{fig:figure2}
\end{figure}

Facial videos and EEG data are recorded simultaneously. EEG is recorded using an ESI NeuroScan System\footnote{http://www.neuroscan.com/} at a sampling rate of 1000 Hz from 62-channel active AgCl electrode cap according to the international 10-20 system. The layout of EEG electrodes on the cap is shown in Fig. \ref{fig:figure2}. The impendence of each electrode has to be less than 5 k$\Omega$. The frontal face videos are recorded from the camera mounted in front of the subjects. Facial videos are encoded into AVI format with the frame rate of 30 frames per second and the resolution of $160\times120$.

\begin{figure}[!t]
\centering
\centerline{\includegraphics[width=3.2in]{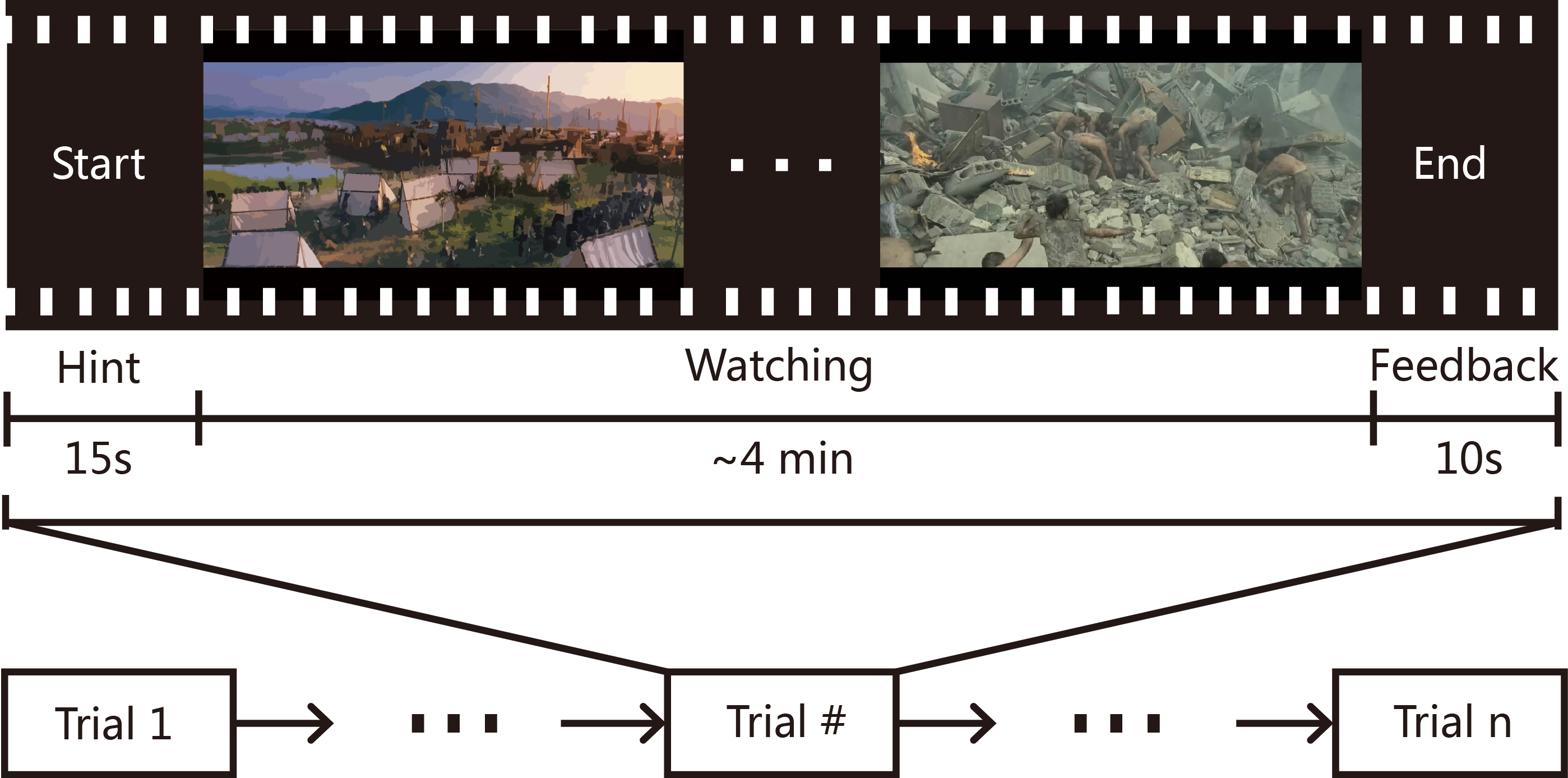}}
\caption{The protocol used in our emotion experiment}\label{fig:figure3}
\end{figure}

There are totally 15 trials for each experiment. There is a 15s hint before each clips and 10s feedback after each clip. For the feedback, participants are told to report their emotional reactions to each film clip by completing the questionnaire immediately after watching each clip. The questions are following Philippot \cite{philippot1993inducing}: (1) what they had actually felt in response to viewing the film clip; (2) how they felt at the specific time they were watching the film clips; (3) have they watched this movie before; (4) have they understood the film clips. They also rate the intensity of subjective emotional arousal using a 5-point scale according to what they actually felt during the task \cite{schaefer2010assessing}. Fig.~\ref{fig:figure3} shows the detailed protocol. For EEG signal processing, the raw EEG data are first downsampled to 200Hz sampling rate. In order to filter the noise and remove the artifacts, the EEG data are then processed with a bandpass filter between 0.5Hz to 70Hz.

\section{Methodology}
\subsection{Feature Extraction}
From our previous work\cite{duan2013differential,zhengeeg}, we have found that the following six different features and electrode combinations are efficient for EEG-based emotion recognition: power spectral density (PSD), differential entropy (DE), differential asymmetry (DASM), rational asymmetry (RASM), asymmetry (ASM) and differential caudality (DCAU) features from EEG. As a result, we use these six different features in this study. According to five frequency bands: delta (1-3Hz); theta (4-7Hz); alpha (8-13Hz); beta (14-30Hz); and gamma (31-50Hz), we compute the traditional PSD features using Short Time Fourier Transform (STFT) with a 1s long window and no overlapping Hanning window. The differential entropy feature is defined as follows \cite{duan2013differential},
\begin{equation}
\begin{split}
h(X)=&-\int_{-\infty}^{\infty}\frac{1}{\sqrt{2\pi\sigma^2}}\exp{\frac{(x-\mu)^2}{2\sigma^2}}\log{\frac{1}{\sqrt{2\pi\sigma^2}}}\\
&\exp{\frac{(x-\mu)^2}{2\sigma^2}}dx=\frac{1}{2}\log{2\pi e\sigma^2},
\end{split}
\end{equation}
where $X$ submits the Gauss distribution $N(\mu,\sigma^2)$, $x$ is a variable, and $\pi$ and $e$ are constants. According to \cite{duan2013differential}, in a certain band, DE is equivalent to the logarithmic power spectral density for a fixed length EEG sequence.


Because many evidences show that the lateralization between the left and right hemisphere is associated with emotions \cite{davidson1992anterior}, we investigate asymmetry features. We compute differential asymmetry (DASM) and rational asymmetry (RASM) features as the differences and ratios between the DE features of 27 pairs of hemispheric asymmetry electrodes (Fp1-Fp2, F7-F8, F3-F4, FT7-FT8, FC3-FC4, T7-T8, P7-P8, C3-C4, TP7-TP8, CP3-CP4, P3-P4, O1-O2, AF3-AF4, F5-F6, F7-F8, FC5-FC6, FC1-FC2, C5-C6, C1-C2, CP5-CP6, CP1-CP2, P5-P6, P1-P2, PO7-PO8, PO5-PO6, PO3-PO4, and CB1-CB2) \cite{duan2013differential}. DASM and RASM can be expressed, respectively, as
\begin{equation}
DASM=DE(X_{left})-DE(X_{right})
\end{equation}
and
\begin{equation}
RASM=DE(X_{left})/DE(X_{right}).
\end{equation}
ASM features are the direct concatenation of DASM and RASM features for comparison. In the literature, the patterns of spectral differences along frontal and posterior brain regions have also been explored \cite{lin2014fusion}. To characterize the spectral-band asymmetry in respect of caudality (in frontal-posterior direction), we define DCAU features as the differences between DE features of 23 pairs of frontal-posterior electrodes (FT7-TP7, FC5-CP5, FC3-CP3, FC1-CP1, FCZ-CPZ, FC2-CP2, FC4-CP4, FC6-CP6, FT8-TP8, F7-P7, F5-P5, F3-P3, F1-P1, FZ-PZ, F2-P2, F4-P4, F6-P6, F8-P8, FP1-O1, FP2-O2, FPZ-OZ, AF3-CB1, and AF4-CB2). DCAU is defined as
\begin{equation}
DCAU = DE(X_{frontal})-DE(X_{posterior}).
\end{equation}
The dimensions of PSD, DE, DASM, RASM, ASM and DCAU are 310 (62 electrodes $\times$ 5 bands), 310 (62 electrodes $\times$ 5 bands), 135 (27 electrode pairs $\times$ 5 bands), 135 (27 electrode pairs $\times$ 5 bands), 270 (54 electrode pairs $\times$ 5 bands), and 115 (23 electrode pairs $\times$ 5 bands), respectively.

\subsection{Feature Smoothing}
Most of the existing approaches to emotion recognition from EEG may be suboptimal because they map EEG signals to static discrete emotional states and do not take temporal dynamics of the emotional state into account. However, in general, emotion should not be considered as a discrete psychophysiological variable \cite{fontaine2007world}. Here, we assume that the emotional state is defined in a continuous space and emotional states change gradually. Our approach focuses on tracking the change of the emotional state over time from EEG. In our approach, we introduce the dynamic characteristics of emotional changes into emotion recognition and investigate how observed EEG is generated from a hidden emotional state. We apply the linear dynamic system (LDS) approach to filter out components which are not associated with emotional states \cite{shi2010off,duan2012eeg}. For comparison, we also evaluate the performance of conventional moving average method.

To make use of the time dependency of emotion changes and further reduce the influence of emotion-unrelated EEG, we introduce the LDS approach to smooth features. A linear dynamic system can be expressed as follows,
\begin{equation}
x_t=z_t+w_t,
\end{equation}
\begin{equation}
z_t=Az_{t-1}+v_t,
\end{equation}
where $x_t$ denotes observed variables, $z_t$ denotes the hidden emotion variables, $A$ is a transition matrix, $w_t$ is a Gaussian noise with mean $\bar w$ and variable $Q$, and $v_t$ is a Gaussian noise with mean $\bar v$ and variable $R$. These equations can also be expressed in an equivalent form in terms of Gaussian conditional distributions,

\begin{equation}
p(x_t|z_t)=N(x_t|z_t+\bar w,Q),
\end{equation}
and
\begin{equation}
p(z_t|z_{t-1})=N(z_t|Az_{t-1}+\bar v,R).
\end{equation}
The initial state is assumed to be
\begin{equation}
p(z_1)=N(z_1|\pi_0,S_0).
\end{equation}
The above model is parameterized by $\theta=\{A,Q,R,\bar w,\bar v,\pi_0,S_0\}$. $\theta$ can be determined using maximum likelihood through the EM algorithm \cite{shumway2010time} based on the observation sequence ${x_t}$. To inference the latent states ${z_t}$ from the observation sequence ${x_t}$, the marginal distribution, $p(z_t|X)$, must be calculated. The latent state can be expressed as
\begin{equation}
z_t=E(z_t|X),
\end{equation}
where $E$ means expectation. This marginal distribution can be achieved by using the messages propagation method \cite{shumway2010time}. We use cross validation to estimate the prior parameters.

\subsection{Dimensionality Reduction}
We compute the initial EEG features based on the signal analysis. However, the features extracted may be uncorrelated with emotion states, and lead to performance degradation of classifiers. What's more, the high dimensionality of features may make classifiers suffer from the `curse of dimensionality' \cite{jain2000statistical}. For real-world applications, dimensionality reduction could help to increase the speed and stability of the classifier. Hence, in this study, we compare two popular approaches: principal component analysis (PCA) algorithm and minimal redundancy maximal relevance (MRMR) algorithm \cite{peng2005feature}.

Principal component analysis (PCA) algorithm uses orthogonal transformation to project high-dimension data to a low-dimension space with a minimal loss of information. This transformation is defined in such a way that the first principal component has the largest possible variance, and each succeeding component in turn has the highest variance possible under the constraint that it is orthogonal to the preceding components.

Although PCA can reduce the feature dimension, it cannot preserve the original domain information such as channel and frequency after the transformation. Hence, we also choose the minimal redundancy maximal relevance (MRMR) algorithm to select a feature subset from an initial feature set. The MRMR algorithm uses mutual information as relevance measure with max-dependency criterion and minimal redundancy criterion. Max-Relevance is to search features satisfying (\ref{maxeq}) with the mean value of all mutual information values between individual feature $x_d$ and class $c$:
\begin{equation}\label{maxeq}
\max D(S,c), D=\frac{1}{|S|}\sum_{x_d\in{S}}{I(x_d;c)},
\end{equation}
where $S$ represents the feature subset to select. When two features highly depend on each other, the respective class-discriminative power would not change much if one of them is removed. Therefore, the following minimal redundancy condition can be added to select mutually exclusive features,
\begin{equation}
\min R(S), R=\frac{1}{|S|^2}\sum_{x_{di},x_{dj}\in{S}}{I(x_{di},x_{dj})}.
\end{equation}
The criterion, combining the above two constraints, is minimal-redundancy-maximal-relevance (MRMR), which can be expressed as
\begin{equation}
\max \varphi(D,R), \varphi=D-R.
\end{equation}
In practice, an incremental search method is used to find the near optimal $K$ features.

\subsection{Classification}
The extracted features are further fed to three conventional pattern classifiers, i.e., \emph{k} nearest neighbors (KNN) \cite{cover1967nearest}, logistic regression (LR) \cite{hosmer2004applied}, and support vector machine (SVM) \cite{vapnik2000nature} and a newly developed pattern classifier, discriminative Graph regularized Extreme Learning Machine (GELM) \cite{ypeng}, to build emotion recognition systems. For the KNN classifier, the Euclidean distance is selected as the distance metric and the number of nearest neighbors is set to 5 using cross validation. For LR, the parameters are computed by maximal likelihood estimation. We use LIBLINEAR software \cite{fan2008liblinear} to build the SVM classifier with linear kernel. The soft margin parameter is selected using cross-validation.

Extreme Learning Machine (ELM) is a single hidden layer feed forward neural network (SLFN) \cite{huang2006extreme}, while learning with local consistency of data has drawn much attention to improve the performance of the existing machine learning models in recent years. Peng \textit{et al.} \cite{ypeng} proposed a discriminative Graph regularized Extreme Learning Machine (GELM) based on the idea that similar samples should share similar properties. GELM obtains much better performance in comparison with other models for face recognition \cite{ypeng} and emotion classification \cite{zhengeeg}.

Given a training data set,
\begin{equation}
L=\{(x_{i},t_{i})|x_{i}\in R^{d},t_{i}\in R^{m}\},
\end{equation}
where $x_{i}=(x_{i1},x_{i2},\cdots, x_{id})^{T}$ and $t_{i}=(t_{i1},t_{i2},\cdot \cdot \cdot, x_{im})^{T}$. In GELM, the adjacent $W$ is defined as follows,

\begin{equation}
x_i=\left\{ \begin{array}{ll} 1/N_{t}, & \textrm{if $h_{i}$ and $h_{j}$ belong to the $t$th class}\\
0, & \textrm{otherwise;}\end{array}\right.
\end{equation}
where $h_{i}=(g_{1}(x_{i}),\cdots,g_{K}(x_{i}))^{T}$ and $h_{j}=(g_{1}(x_{j}),\cdots,g_{K}(x_{j}))^{T}$ are hidden layer outputs for two input samples $x_{i}$ and $x_{j}$ . Then we can compute the graph Laplacian $L=D-W$ , where $D$ is a diagonal matrix  and each of the entries in $D$ is the column sums of $W$. Therefore, GELM can incorporate two regularization terms into conventional ELM model. The objective function of GELM is defined as follows,

\begin{equation}\label{10}
\min_{\beta}||H\beta-T||_{2}^{2}+\lambda_{1}Tr(H\beta L \beta^{T}H^{T})+\lambda_{2}||\beta||_{2}^{2}
\end{equation}
where $Tr(H\beta L \beta^{T}H^{T})$ is the graph regularization term, $||\beta||^{2}$ is the $l_{2}$-norm regularization term, and $\lambda_{1}$ and $\lambda_{2}$ are regularization parameters to balance two terms.

By setting the differentiate of the objective function (\ref{10}) with respect to $\beta$ as zero, we have

\begin{equation}\label{11}
\beta = (HH^{T}+\lambda_{1}HLH^{T}+\lambda_{2}I)^{-1}HT
\end{equation}

In GELM, the constraint imposed on output weights enforces the outputs of samples from the same class to be similar. The constraint can be formulated as a regularization term to the objective function of basic ELM, which also makes the output weight matrix calculated directly.

\section{Experiment Results}
\subsection{Experiment Results on DEAP Data}
In this section, to validate the efficiency of the machine learning algorithms used in this study, we first evaluate these algorithms with the publicly available emotion dataset, the DEAP dataset \footnote{http://www.eecs.qmul.ac.uk/mmv/datasets/deap/index.html} \cite{koelstra2012deap} and compare the performance of our models with other methods used in the existing studies on the same emotion EEG dataset.

The DEAP dataset consists of EEG and peripheral physiological signals of 32 participants as each watched 40 excerpts of one-minute duration music videos. The EEG signals were recorded from 32 active electrodes (channels) according to the international 10-20 system, whereas peripheral physiological signals (8 channels) include galvanic skin response, skin temperature, blood volume pressure, respiration rate, electromyogram and electrooculogram (horizontal and vertical). The participants rated each video in terms of the levels of arousal, valence, like/dislike, dominance, and familiarity. More details about the DEAP dataset are given in \cite{koelstra2012deap}.



\begin{figure}[!b]
\centering
\centerline{\includegraphics[width=3.5in]{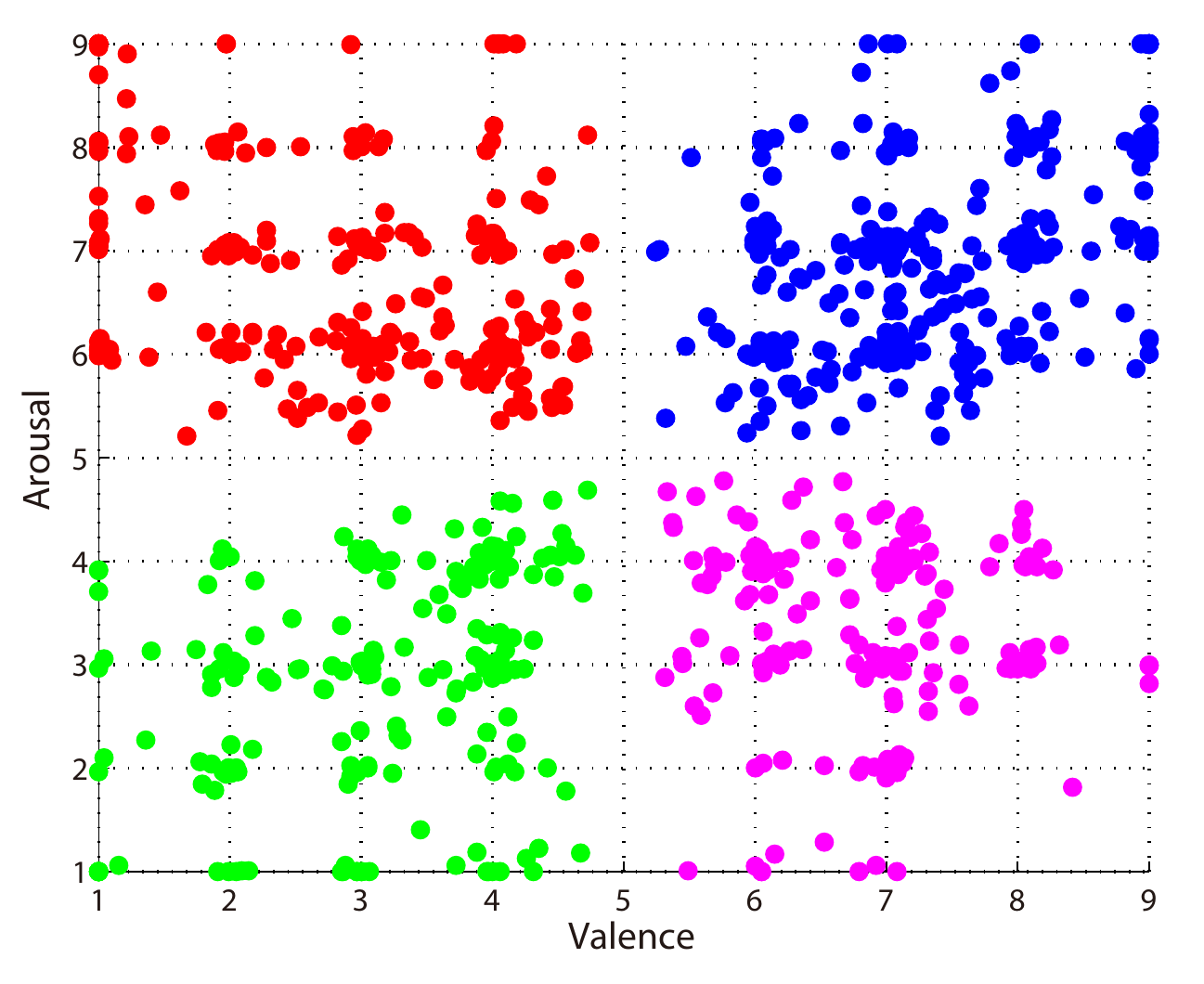}}
\caption{The rating distribution of DEAP on the arousal-valence plane (VA plane) for the four conditions (LALV, HALV, LAHV, and HAHV)}\label{fig:figure5}
\end{figure}

In this experiment, we adopt an emotion representation model based on the valence-arousal model. Each dimension has value ranging from 1 to 9. Further, we segment the four quadrants of the valence-arousal (VA) space according to the ratings. LALV, HALV, LAHV, and HAHV denote low arousal/low valence, high arousal/low valence, low arousal/high valence, and high arousal/high valence, respectively. Considering the fuzzy boundary of emotions and the variations of participants' ratings possibly associated with individual difference in rating scale, we add an gap to segment the quadrants of VA space to ensure the correct ratings of participants' true self-elicitation emotion and discard the EEG data whose ratings of arousal and valence are between 4.8 and 5.2. The numbers of instances for LALV, HALV, LAHV, and HAHV are 12474, 16128, 10962 and 21420, respectively. The rating distribution of DEAP on the arousal-valence plane (VA plane) for the four conditions is shown in Fig. \ref{fig:figure5}. We can see that the ratings are distributed approximately uniformly \cite{koelstra2012deap}. We label the EEG data according to the participants' ratings of valence and arousal.

We firstly extracted the PSD, DE, DASM, RASM, ASM and DCAU features of the 32-channel EEG data. The original EEG data that we got from DEAP dataset are preprocessed with a downsample to 128 Hz and a bandpass frequency filter from 4.0-45.0Hz and EOG artifacts are removed. So we extract the features in the four frequency bands: theta: 4-7Hz, alpha: 8-13Hz, beta: 14-30Hz, and gamma: 31-45Hz. The features are further smoothed with the linear dynamic system approach. Then we select SVM and GELM as classifiers. In this study, we use SVM classifier with linear kernel and the number of hidden layer neurons for GELM is fixed as 10 times of the dimensions of input. To use the entire data set for training and testing the classifiers, a 5-fold cross-validation scheme is adopted. All experiments here are performed with 5-fold cross-validation and classification performance is evaluated through the classification accuracy rate.


\begin{table}[h]
\caption{The mean accuracy rates (\%) of SVM and GELM classifiers for different features obtained from separate and total frequency bands.}\label{tab:table4}
\renewcommand{\arraystretch}{1.3}
\begin{center}
\begin{tabular}{p{2.2em}lp{2.em}llllll}
\hline
\hline
Feature               & Classifier & Theta & Alpha & Beta  & Gamma & Total \\ \hline
\multirow{2}{*}{PSD}  & SVM        & 32.86 & 33.49 & 33.73 & 31.99 & \textbf{36.19} \\
                      & GELM       & \textbf{61.78} & 61.14 & 61.77 & 61.56 & 61.46 \\ \hline
\multirow{2}{*}{DE}   & SVM        & 44.31 & 41.59 & 43.54 & 42.74 & \textbf{47.57} \\
                      & GELM       & 61.45 & 61.65 & 62.17 & 61.45 & \textbf{69.67} \\ \hline
\multirow{2}{*}{DASM} & SVM        & \textbf{43.18} & 42.72 & 42.07 & 41.24 & 40.70 \\
                      & GELM       & \textbf{57.86} & 57.02 & 56.08 & 56.48 & 52.54 \\ \hline
\multirow{2}{*}{RASM} & SVM        & \textbf{54.34} & 52.54 & 53.12 & 52.76 & 51.83 \\
                      & GELM       & 46.66 & 44.52 & 44.88 & 45.56 & \textbf{52.70} \\ \hline
\multirow{2}{*}{ASM}  & SVM        & \textbf{45.03} & 44.03 & 43.59 & 44.03 & 40.76 \\
                      & GELM       & \textbf{56.16} & 54.30 & 54.40 & 54.73 & 51.82 \\ \hline
\multirow{2}{*}{DCAU} & SVM        & \textbf{41.51} & 41.05 & 41.27 & 40.00 & 40.85 \\
                      & GELM       & 57.47 & 56.58 & \textbf{58.25} & \textbf{58.25} & 55.26 \\ \hline
                      \hline
\end{tabular}
\end{center}
\end{table}


Table \ref{tab:table4} shows the mean accuracy rates of SVM and GELM classifiers for different features obtained from various frequency bands (theta, alpha, beta and gamma) and total frequency bands. It should be noted that `Total' in Table \ref{tab:table4} represents the direct concatenation of all features from four frequency bands. Since the EEG data of DEAP are preprocessed with a bandpass frequency filter from 4.0-45.0Hz, the results of delta frequency bands are not included. The average accuracies (\%) are 61.46, 69.67, 52.54, 52.70, 51.82 and 55.26 for PSD, DE, DASM, RASM, ASM and DCAU features from the total frequency bands, respectively. The best accuracy of GELM classifier is 69.67\% using DE features of total frequency bands while the best accuracy of SVM classifier is 54.34\%. We also evaluate the performance of KNN, logistic regression and SVM with RBF kernel on DEAP, which achieve the accuracies (\%) and standard deviations (\%) of 35.50/14.50, 40.86/16.87, and 39.21/15.87, respectively, using the DE features of total frequency bands. We perform one way analysis of variance (ANOVA) to study the statistical significance. The DE features outperform the PSD features significantly ($p<0.01$) and for classifiers, GELM performs much better than SVM ($p<0.01$). As we can also see from Table \ref{tab:table4}, the diversity of classification accuracy for different frequency bands is not significant for the DEAP dataset ($p>0.95$). The results here do not show specific frequency bands for the quadrants of VA space. We can also see that the DCAU features achieve comparable accuracies. These results indicate that there exists some kind asymmetry which has discriminative information for the four affect elicitation conditions (LALV, HALV, LAHV, and HAHV), as discussed in Section 2.2.

\begin{table}[h]\small
\caption{Comparison of various studies using EEG in the DEAP dataset. Our method achieves the best performance among these approaches. }\label{tab:table5}
\renewcommand{\arraystretch}{1.3}
\begin{center}
\begin{tabular}{p{1.5cm}|p{6cm}}
\hline
\hline
Study                    & Results                                                                          \\ \hline
Chung \emph{et al.} \cite{chung2012affective}    & 66.6\%, 66.4\% for valence and arousal (2 classes), 53.4\%, 51.0\% for valence and arousal (3 classes) with all 32 participants.  \\ \hline
Koelstra \emph{et al.} \cite{koelstra2012deap} & 62.0\%, 57.6\% for valence and arousal (2 classes) with all 32 participants.             \\ \hline
Liu \emph{et al.} \cite{liu2013real}      & 63.04\% for arousal-dominance recognition (4 classes) with selected 10 participants.                     \\ \hline
Zhang \emph{et al.} \cite{zhang2013ontology}    & 75.19 \% and 81.74 \% on valence and arousal (2 classes) with selected eight participants. \\ \hline
Our method               & 69.67\% for quadrants of VA space (4 classes) with all 32 participants.                                 \\ \hline
\hline
\end{tabular}
\end{center}
\end{table}

The recognition accuracy comparison of various systems using EEG signals in DEAP dataset is presented in Table \ref{tab:table5}. Here single modality signal (EEG) is used rather than in a combined modality fusion way. Chung \emph{et al.} \cite{chung2012affective} defined a weighted-log-posterior function for the Bayes classifier and evaluated the method with DEAP dataset. Their accuracies for valence and arousal classification are 66.6\% and 66.4\% for two classes and 53.4\% and 51.0\% for three classes, respectively. Koelstra \emph{et al.} \cite{koelstra2012deap} developed the DEAP dataset and they obtained an average accuracy of 62.0\%, 57.6\% for valence and arousal (2 classes), respectively. Liu \emph{et al.} \cite{liu2013real} proposed a real-time fractal dimension (FD) based valence level recognition algorithm from EEG signals and got a mean accuracy of 63.04\% for arousal-dominance recognition (4 classes) with selected 10 participants. Zhang \emph{et al.} \cite{zhang2013ontology} described an ontological model for representation and integration of EEG data and their model reached an average recognition ratio of 75.19\% on valence and 81.74\% on arousal for eight participants. Although their accuracies were relatively high, the categories of each dimension are only two and these results were achieved with a subset of the original dataset. In contrast, our method achieves an average accuracy of 69.67\% on the same data set for quadrants of VA space (LALV, HALV, LAHV, and HAHV) with PSD features from theta frequency band for all 32 participants.

\subsection{Experiment Results on SEED data}

In this section, we present the results of our approaches evaluated with our SEED dataset. One very important difference between SEED and DEAP is that SEED contains three sessions at the time interval of one week or longer for the same subject.

\subsubsection{Performance of Emotion Recognition Models}
We first compare six different features, namely PSD, DE, DASM, RASM, ASM and DCAU, from total frequency bands. Here we use GELM as classifier and the number of hidden layer neurons is fixed as 10 times of the dimensions of input. We adopt a 5-fold cross-validation scheme. From Table \ref{tab:tablenew}, we can see that DE features have the highest accuracy and the lowest standard deviation than traditional PSD features, which imply that DE features are more suitable for EEG-based emotion recognition than other five different features. For the asymmetry features, although they have fewer dimensions than PSD features, they can achieve significant better performance than PSD, which means that brain processing about positive, neutral and negative emotion has asymmetrical characteristics.

\begin{table}[!h]
\caption{The means and standard deviations of accuracies in percentage (\%) for PSD, DE, DASM, RASM, ASM and DCAU features from total frequency bands.}\label{tab:tablenew}
\begin{center}
\begin{tabular}{ccccccc}
\hline
\hline
Feature & PSD         & DE            & DASM & RASM & ASM & DCAU            \\ \hline
Mean   & 72.75 & \textbf{91.07} & 86.76 &86.98 &85.27 &89.95\\ \hline
Std.   & 11.85 & \textbf{7.54}  & 8.43 &9.09 &9.16 &6.87 \\ \hline
\hline
\end{tabular}
\end{center}
\end{table}



\begin{figure}[!b]
\centerline{\includegraphics[width=3.4in]{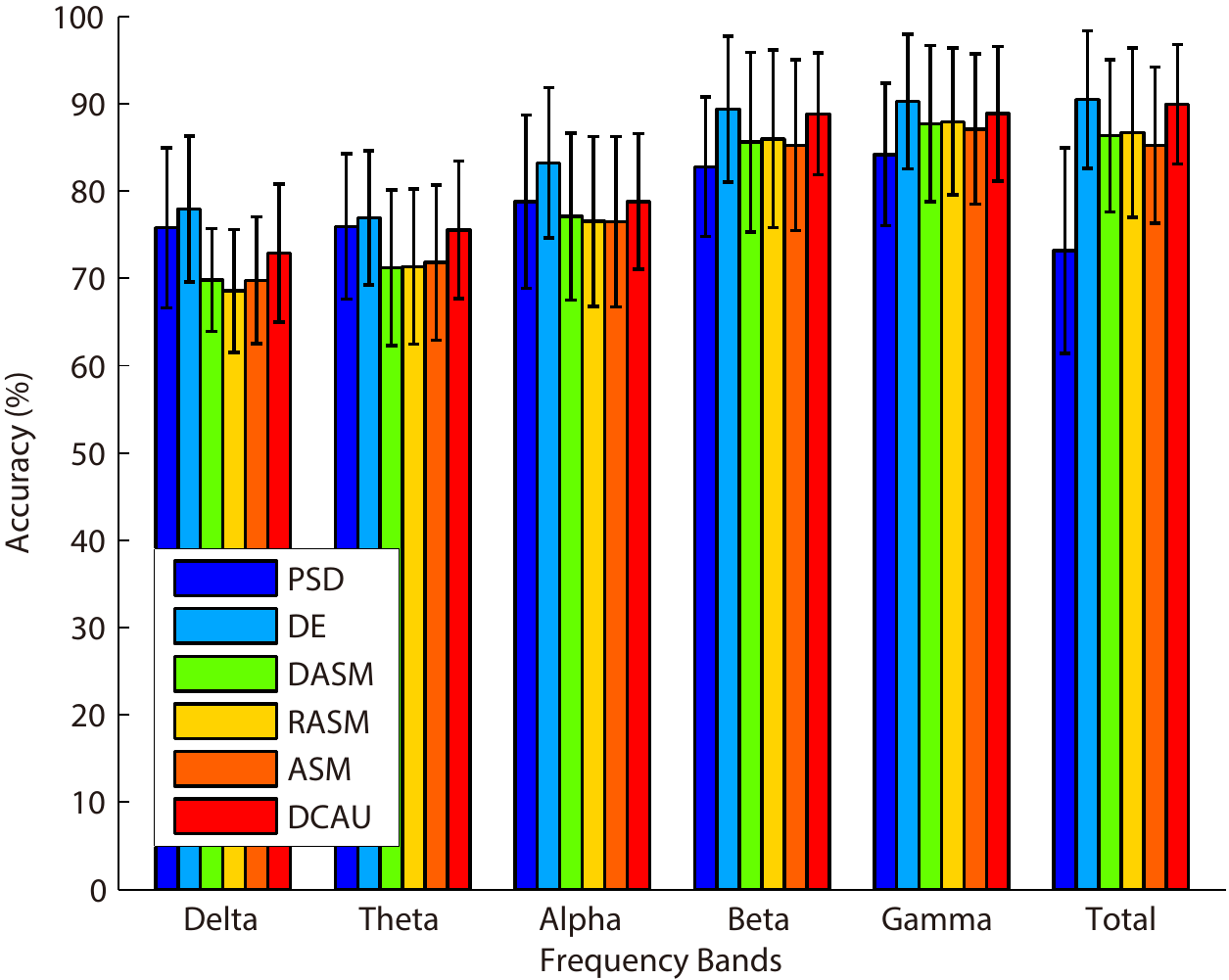}}
\caption{The average accuracies of GELM using different features obtained from five frequency bands and in a fusion method.}\label{fig:figure7}
\end{figure}

\begin{figure}[!b]
\centering
\centerline{\includegraphics[width=3.2in]{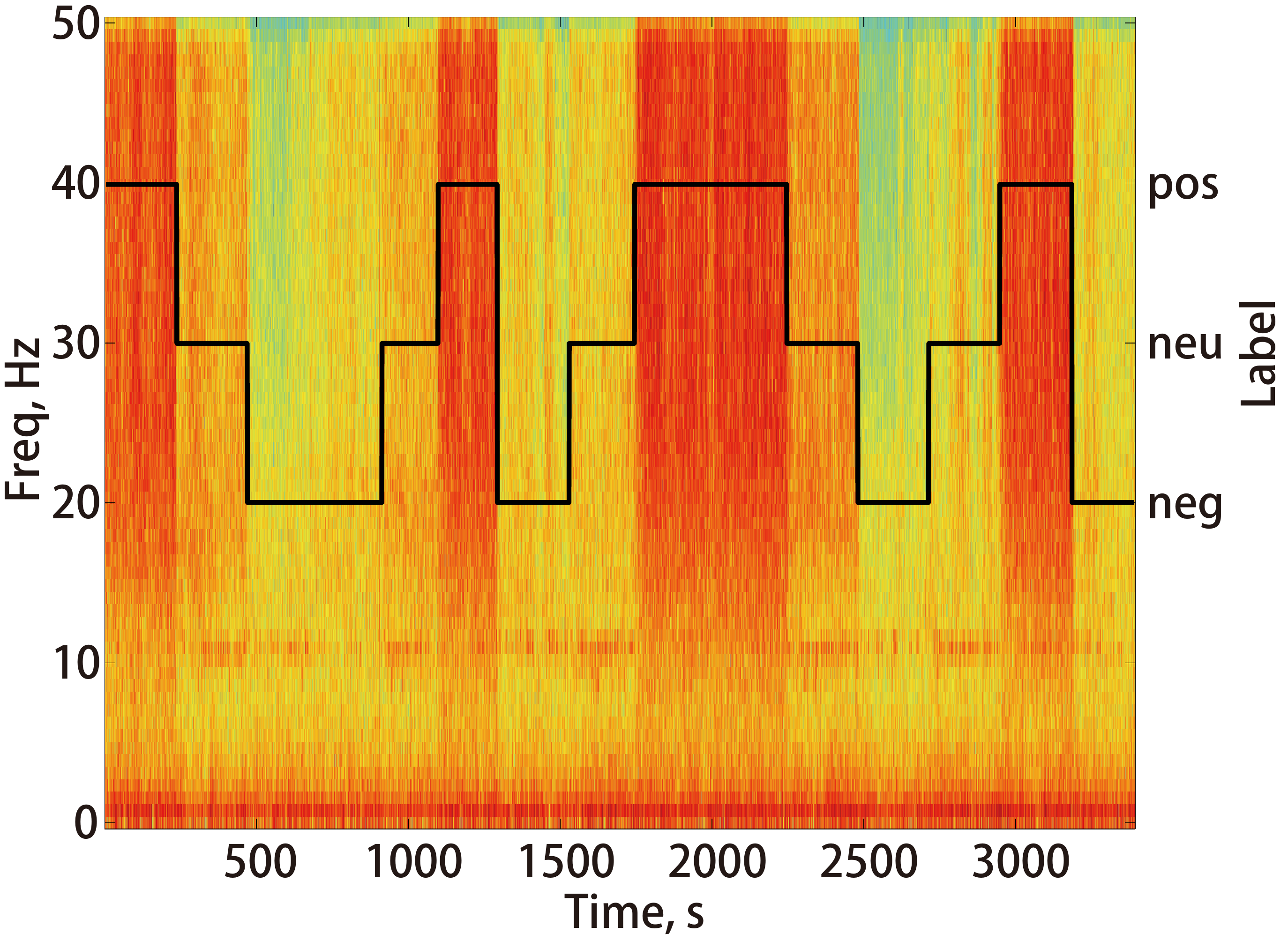}}
\caption{The spectrogram of one electrode position T7 in one experiment. As different emotions elicited, we can see that the spectrogram has different patterns.}\label{fig:figure8}
\end{figure}

We also evaluate the performance of two different feature smoothing algorithms. Here, we compare the linear dynamic system (LDS) approach with the conventional moving average algorithm. The size of moving windows is five in this study.  The means and standard deviations of accuracies in percentage (\%) for without smoothing, moving average, and the LDS approach are 70.82/9.17, 76.07/8.86 and 91.07/7.54, respectively. We can see that the LDS approach significantly outperforms the moving average method ($p<0.01$), which achieves 14.41\% higher for accuracy. And the results also demonstrate that feature smoothing plays a significant role in EEG-based emotion recognition.


We compare the performance of four different classifiers, KNN, Logistic Regression, SVM and GELM. In this evaluation, DE features of 310 dimensions were used as the inputs of classifiers. The parameter $K$ of KNN was fixed constant five. For LR and linear SVM, grid search with cross-validation was used to tune the parameters. The mean accuracies and standard deviations in percentage (\%) of KNN, LR, SVM with RBF kernel, SVM with linear kernel and GELM are 70.43/12.73, 84.08/8.77, 78.21/9.72, 83.26/9.08 and 91.07/7.54, respectively. From the above results, we can see that GELM outperforms other classifiers with higher accuracies and lower standard deviations, which imply that GELM is more suited for EEG-based emotion recognition. In GELM, the constraint imposed on output weights enforces the output of samples from the same class to be similar.

\subsubsection{Neural Signatures and Stable Patterns}

\begin{figure*}[!htbp]
\centering
\centerline{\includegraphics[width=6.8in]{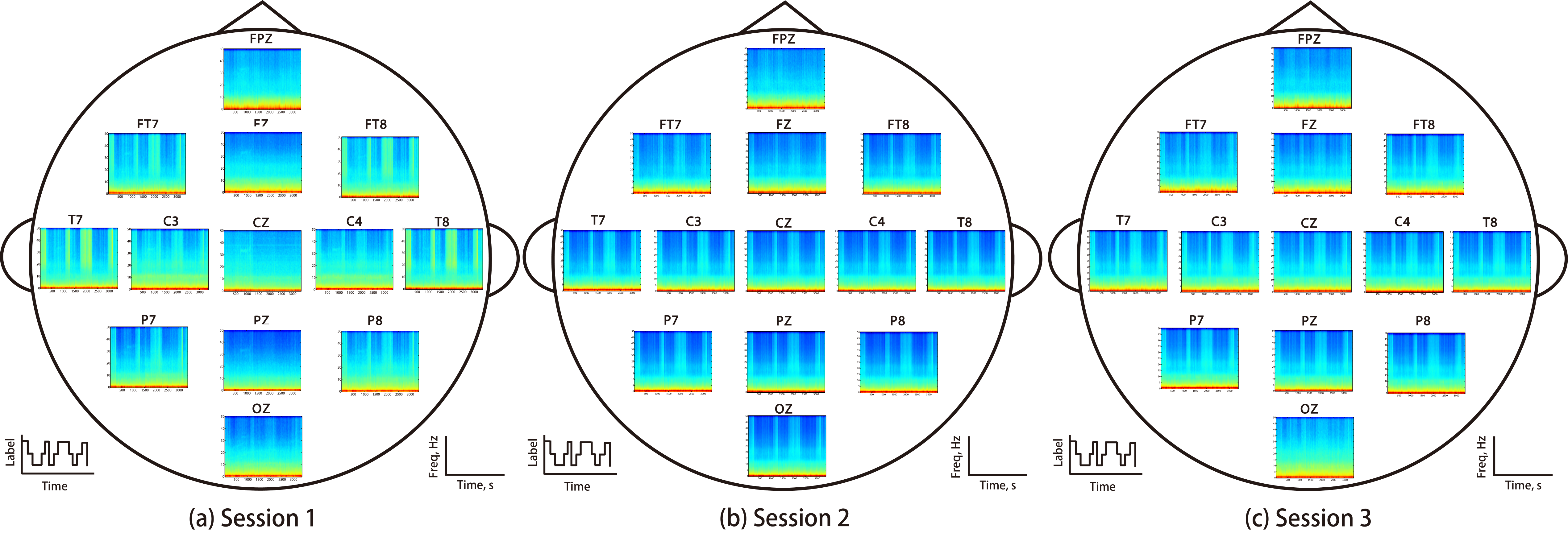}}
\caption{The average spectrogram over subjects for each session at some electrodes, which shows the stable neural patterns over time in temporal lobes and high frequency bands. (Red color indicates high amplitude.)}\label{fig:figure8_new}
\end{figure*}

\begin{figure*}[!htbp]
\centering
\centerline{\includegraphics[width=6.8in]{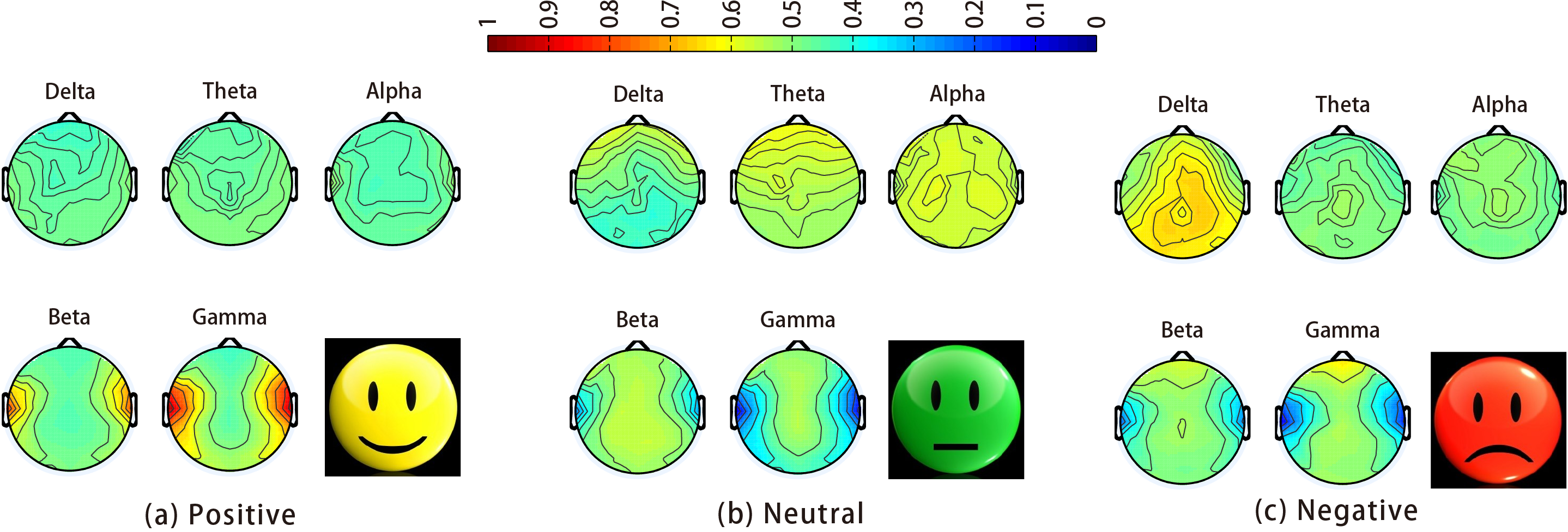}}
\caption{The average neural patterns over all subjects and sessions for different emotions, which shows that neural signatures associated with positive, neutral and negative emotions do exist. The lateral temporal areas activate more for positive emotion than negative one in beta and gamma bands. While the neural patterns of neutral emotion are similar to that of negative emotion, which both have less activation in temporal areas, the neural patterns of neutral emotion have higher alpha responses at parietal and occipital sites. The negative emotion patterns have significant higher delta responses at parietal and occipital sites and higher gamma responses at prefrontal sites.}\label{fig:figure9}
\end{figure*}

Fig. \ref{fig:figure7} presents the average accuracies of GELM classifier with the six different features extracted from five frequency bands (Delta, Theta, Alpha, Beta and Gamma) and the direct concatenation of these five frequency bands. The results in Fig. \ref{fig:figure7} indicate that features obtained from gamma and beta frequency bands perform better than other frequency bands, which imply that beta and gamma oscillation of brain activity are more related with the processing of these three emotional states than other frequency oscillation as described in \cite{li2009emotion,guntekin2010event,martini2012dynamics}.

Fig. \ref{fig:figure8} shows the spectrogram of one electrode position T7 in a experiment and Fig. \ref{fig:figure8_new} shows the average spectrogram over subjects for each session at some electrodes (FPZ, FT7, F7, FT8, T7, C3, CZ, C4, T8, P7, PZ, P8 and OZ). As we can see from Figs. \ref{fig:figure8} and \ref{fig:figure8_new}, the spectrograms have different patterns for different elicited emotions. The dynamics of higher frequency oscillations are more related to positive/negative emotions, especially for the temporal lobes. Moreover, the neural patterns over time are relatively stable for each session. To see the neural patterns associated with emotion processing, we project the DE features to the scalp to find temporal dynamics of frequency oscillations and stable patterns across subjects.

Fig. \ref{fig:figure9} depicts the average neural patterns for positive, neutral, and negative emotion. The results demonstrate that neural signatures associated with positive, neutral, and negative emotions do exist. The lateral temporal areas activate more for positive emotion than negative one in beta and gamma bands, while the energy of the prefrontal area enhances for negative emotion over positive one in beta and gamma bands. While the neural patterns of neutral emotion are similar to negative emotion, which both have less activation in temporal areas, the neural patterns of neutral emotion have higher alpha responses at parietal and occipital sites. For negative emotion, the neural patterns have significant higher delta responses at parietal and occipital sites and significant higher gamma responses at prefrontal sites. The existing studies \cite{ray1985eeg}, \cite{klimesch1998induced} have shown that EEG alpha activity reflects attentional processing and beta activity reflects emotional and cognitive processes. When participants watched neutral stimuli, they tended to be more relaxed and less attentional, which evoked alpha responses. And when processing positive emotion processing, the energy of beta and gamma response enhanced. Our results are consistent with the findings of the existing work \cite{hadjidimitriou2012toward,jenke2014feature,ray1985eeg}.

\subsubsection{Dimensionality Reduction}

As discussed above, the brain activities of emotion processing have critical frequency bands and brain areas, which imply that there must be a low-dimension manifold structure for emotion related EEG signals. Therefore, we investigate how the dimension of features will affect the performance of emotion recognition. Here, we compare two dimensionality reduction algorithms, the principle component analysis (PCA) algorithm and the minimal redundancy maximal relevance (MRMR) algorithm, with DE features of 310 dimensions as inputs and GELM as a classifier.

\begin{figure}[!b]
\centering
\centerline{\includegraphics[width=3in]{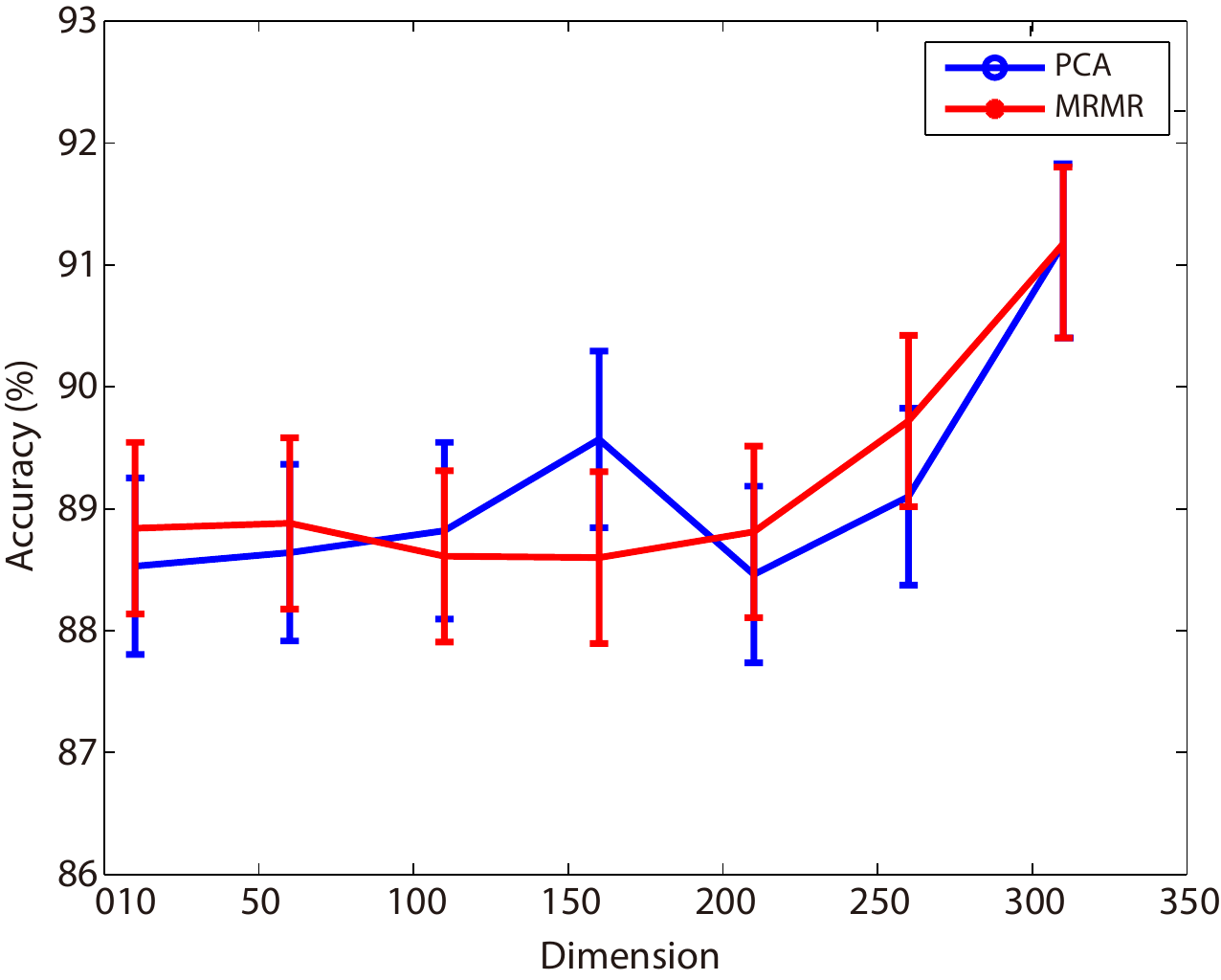}}
\caption{The results of dimensionality reduction using PCA and MRMR}\label{fig:figure10}
\end{figure}

Fig. \ref{fig:figure10} shows the results of dimensionality reduction. We find that dimensionality reduction does not affect the performance of our model very much. For PCA algorithm, when the dimension is reduced to 210, the accuracy drops from 91.07\% to 88.46\% and then reached a local maximum value 89.57\% at dimension of 160. For MRMR algorithms, the accuracies vary slightly with lower dimension features. 

From the performance comparison between PCA and MRMR, we see that it is better to apply MRMR algorithm in EEG-based emotion recognition. Because MRMR algorithm finds the best emotion-relevant and minimal redundancy features. It also preserves original domain information such as channel and frequency bands, which have most discriminative information for emotion recognition after the transformation. This discovery helps us to reduce the computations of features and the complexity of the computational models.

\begin{figure}[!h]
\centering
\centerline{\includegraphics[width=\linewidth]{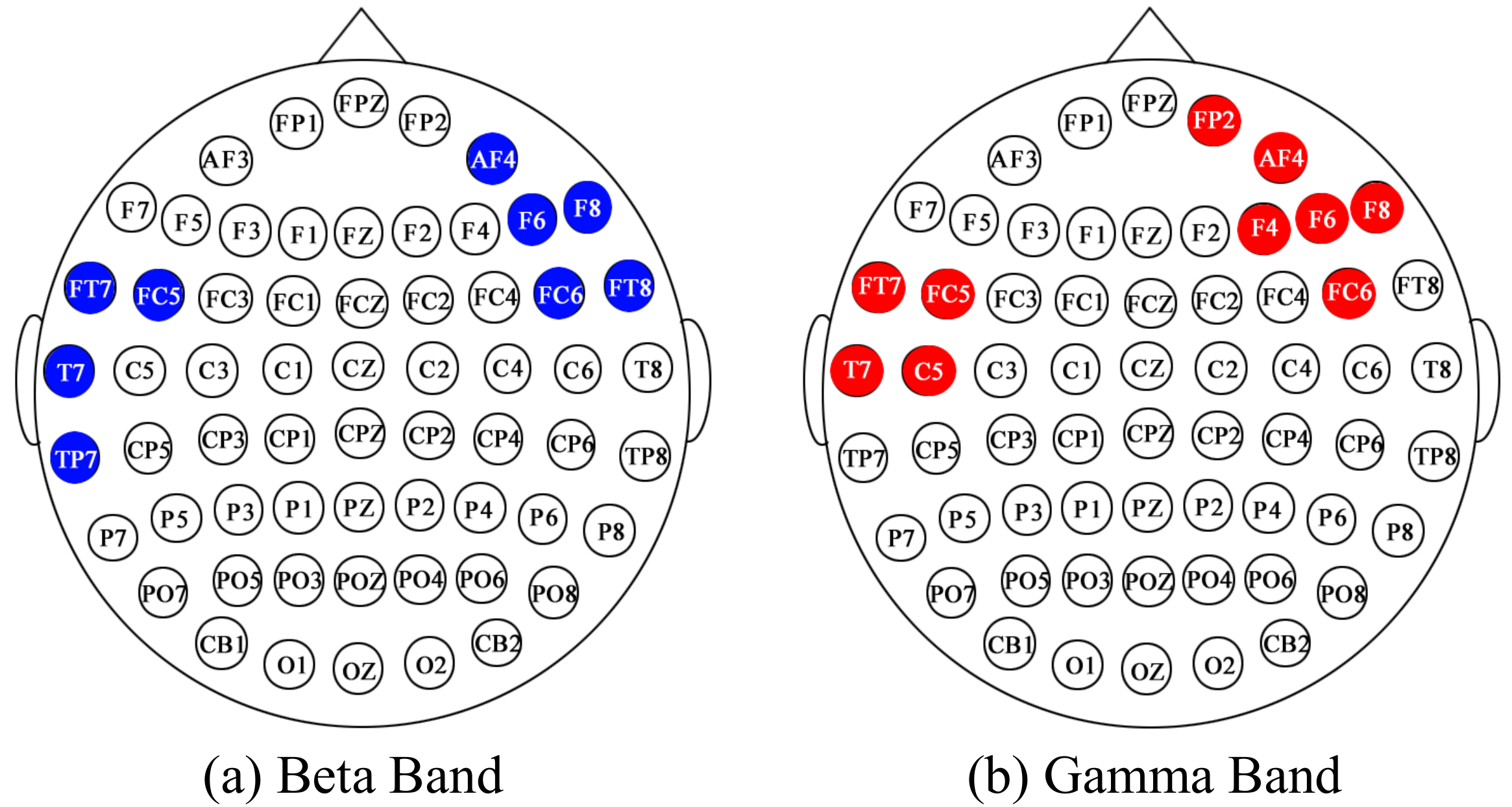}}
\begin{flushleft}
\footnotesize{One electrode location ``FT8" is from alpha frequency band}
\end{flushleft}
\caption{Distribution of the top 20 subject-independent features selected by correlation coefficient.}\label{fig:figure11}
\end{figure}

Fig. \ref{fig:figure11} presents the distribution of the top 20 subject-independent features selected by correlation coefficient. The top 20 features were selected from alpha frequency bands at electrode locations (FT8), beta frequency bands at electrode locations (AF4, F6, F8, FT7, FC5, FC6, FT8, T7, TP7) and gamma frequency band at electrode locations (FP2, AF4, F4, F6, F8, FT7, FC5, FC6, T7, C5). These selected features are mostly from beta and gamma frequency bands and at lateral temporal and frontal brain areas, which is consistent with the above findings of time-frequency analysis.

\subsubsection{Stability of Emotion Recognition Model over Time}

It should be noted that SEED consists of 15 participants and each one performed the experiments three times. The interval of two sessions is one week or longer. This is the novelty of SEED we developed in comparison with the existing emotion EEG datasets. By using SEED, we evaluate whether the performance of our emotion recognition model is stable with the passage of time. We split the data from different sessions of one participant to training data and testing data and trained the model with GELM. The features employed here are the DE features extracted from total frequency bands after LDS smoothing. 

\begin{table*}[!htbp] \normalsize
\caption{The classification accuracies (\%) of training and test data from different sessions using GELM.}\label{tab:table8}
\begin{center}
\begin{tabular}{|c|l|c|c|c||c|c|c|c|c}
\hline
\hline
\multirow{2}{*}{Subject} & \multirow{2}{*}{Train} & \multicolumn{3}{c||}{Test} & \multicolumn{1}{c|}{\multirow{2}{*}{Subject}} & \multicolumn{1}{c|}{\multirow{2}{*}{Train}} & \multicolumn{3}{c|}{Test}                                                               \\ \cline{3-5} \cline{8-10}
                         &                        & 1st   & 2nd & 3rd  & \multicolumn{1}{c|}{}                         & \multicolumn{1}{c|}{}                       & \multicolumn{1}{c|}{1st}  & \multicolumn{1}{c|}{2nd} & \multicolumn{1}{c|}{3rd}  \\ \hline
\multirow{3}{*}{\#1}     & 1st                  & \textbf{91.62}   & 60.26  & 60.26  & \multicolumn{1}{c|}{\multirow{3}{*}{\#2}}     & \multicolumn{1}{l|}{1st}                  & \multicolumn{1}{c|}{\textbf{100.00}} & \multicolumn{1}{c|}{68.28}  & \multicolumn{1}{c|}{60.84}  \\ \cline{2-5} \cline{7-10}
                         & 2nd                 & 68.28   & \textbf{80.2}   & 68.28  & \multicolumn{1}{c|}{}                         & \multicolumn{1}{l|}{2nd}                 & \multicolumn{1}{c|}{\textbf{85.12}}  & \multicolumn{1}{c|}{71.68}  & \multicolumn{1}{c|}{81.65}  \\ \cline{2-5} \cline{7-10}
                         & 3rd                  & 68.28   & 52.53  & \textbf{92.56}  & \multicolumn{1}{c|}{}                         & \multicolumn{1}{l|}{3rd}                  & \multicolumn{1}{c|}{85.12}  & \multicolumn{1}{c|}{80.42}  & \multicolumn{1}{c|}{\textbf{90.82}}  \\ \hline
\multirow{3}{*}{\#3}     & 1st                  & 95.95   & \textbf{100.00} & 75.51  & \multicolumn{1}{c|}{\multirow{3}{*}{\#4}}     & \multicolumn{1}{l|}{1st}                  & \multicolumn{1}{c|}{\textbf{100.00}} & \multicolumn{1}{c|}{68.28}  & \multicolumn{1}{c|}{68.28}  \\ \cline{2-5} \cline{7-10}
                         & 2nd                 & 76.95   & \textbf{97.04}  & 82.95  & \multicolumn{1}{c|}{}                         & \multicolumn{1}{l|}{2nd}                 & \multicolumn{1}{c|}{83.02}  & \multicolumn{1}{c|}{\textbf{100.00}} & \multicolumn{1}{c|}{91.69}  \\ \cline{2-5} \cline{7-10}
                         & 3rd                  & 80.20   & \textbf{88.08}  & 68.93  & \multicolumn{1}{c|}{}                         & \multicolumn{1}{l|}{3rd}                  & \multicolumn{1}{c|}{76.81}  & \multicolumn{1}{c|}{\textbf{100.00}} & \multicolumn{1}{c|}{\textbf{100.00}} \\ \hline
\multirow{3}{*}{\#5}     & 1st                  & \textbf{75.94}   & 59.61  & 61.05  & \multicolumn{1}{c|}{\multirow{3}{*}{\#6}}     & \multicolumn{1}{l|}{1st}                  & \multicolumn{1}{c|}{\textbf{79.70}}  & \multicolumn{1}{c|}{71.60}  & \multicolumn{1}{c|}{53.83}  \\ \cline{2-5} \cline{7-10}
                         & 2nd                 & \textbf{80.78}   & 75.00  & 69.65  & \multicolumn{1}{c|}{}                         & \multicolumn{1}{l|}{2nd}                 & \multicolumn{1}{c|}{67.92}  & \multicolumn{1}{c|}{\textbf{80.78}}  & \multicolumn{1}{c|}{55.71}  \\ \cline{2-5} \cline{7-10}
                         & 3rd                  & 56.50   & 54.48  & \textbf{98.12}  & \multicolumn{1}{c|}{}                         & \multicolumn{1}{l|}{3rd}                  & \multicolumn{1}{c|}{70.30}  & \multicolumn{1}{c|}{71.75}  & \multicolumn{1}{c|}{\textbf{86.27}}  \\ \hline
\multirow{3}{*}{\#7}     & 1st                  & \textbf{90.39}   & 66.69  & 66.47  & \multicolumn{1}{c|}{\multirow{3}{*}{\#8}}     & \multicolumn{1}{l|}{1st}                  & \multicolumn{1}{c|}{\textbf{75.07}}  & \multicolumn{1}{c|}{67.77}  & \multicolumn{1}{c|}{51.01}  \\ \cline{2-5} \cline{7-10}
                         & 2nd                 & 67.49   & \textbf{95.81}  & 59.83  & \multicolumn{1}{c|}{}                         & \multicolumn{1}{l|}{2nd}                 & \multicolumn{1}{c|}{72.83}  & \multicolumn{1}{c|}{\textbf{91.33}}  & \multicolumn{1}{c|}{45.95}  \\ \cline{2-5} \cline{7-10}
                         & 3rd                  & \textbf{83.60}   & 81.79  & 74.93  & \multicolumn{1}{c|}{}                         & \multicolumn{1}{l|}{3rd}                  & \multicolumn{1}{c|}{59.61}  & \multicolumn{1}{c|}{\textbf{75.94}}  & \multicolumn{1}{c|}{73.05}  \\ \hline
\multirow{3}{*}{\#9}     & 1st                  & \textbf{91.98}   & 80.56  & 78.47  & \multicolumn{1}{c|}{\multirow{3}{*}{\#10}}    & \multicolumn{1}{l|}{1st}                  & \multicolumn{1}{c|}{\textbf{85.12}}  & \multicolumn{1}{c|}{70.38}  & \multicolumn{1}{c|}{69.29}  \\ \cline{2-5} \cline{7-10}
                         & 2nd                 & 81.36   & \textbf{100.00} & 95.95  & \multicolumn{1}{c|}{}                         & \multicolumn{1}{l|}{2nd}                 & \multicolumn{1}{c|}{60.12}  & \multicolumn{1}{c|}{86.05}  & \multicolumn{1}{c|}{\textbf{87.14}}  \\ \cline{2-5} \cline{7-10}
                         & 3rd                  & 93.42   & \textbf{95.52}  & 93.42  & \multicolumn{1}{c|}{}                         & \multicolumn{1}{l|}{3rd}                  & \multicolumn{1}{c|}{87.07}  & \multicolumn{1}{c|}{83.02}  & \multicolumn{1}{c|}{\textbf{95.74}}  \\ \hline
\multirow{3}{*}{\#11}    & 1st                  & \textbf{96.24}   & 67.99  & 76.01  & \multicolumn{1}{c|}{\multirow{3}{*}{\#12}}    & \multicolumn{1}{l|}{1st}                  & \multicolumn{1}{c|}{\textbf{86.78}}  & \multicolumn{1}{c|}{74.86}  & \multicolumn{1}{c|}{63.29}  \\ \cline{2-5} \cline{7-10}
                         & 2nd                 & 77.89   & 85.33  & \textbf{95.59}  & \multicolumn{1}{c|}{}                         & \multicolumn{1}{l|}{2nd}                 & \multicolumn{1}{c|}{84.39}  & \multicolumn{1}{c|}{\textbf{91.62}}  & \multicolumn{1}{c|}{68.06}  \\ \cline{2-5} \cline{7-10}
                         & 3rd                  & 65.39   & 66.04  & \textbf{100.00} & \multicolumn{1}{c|}{}                         & \multicolumn{1}{l|}{3rd}                  & \multicolumn{1}{c|}{75.58}  & \multicolumn{1}{c|}{\textbf{83.45}}  & \multicolumn{1}{c|}{73.48}  \\ \hline
\multirow{3}{*}{\#13}    & 1st                  & \textbf{93.71}   & 76.88  & 92.63  & \multicolumn{1}{c|}{\multirow{3}{*}{\#14}}    & \multicolumn{1}{l|}{1st}                  & \multicolumn{1}{c|}{\textbf{100.00}} & \multicolumn{1}{c|}{86.27}  & \multicolumn{1}{c|}{64.02}  \\ \cline{2-5} \cline{7-10}
                         & 2nd                 & 70.38   & 90.75  & \textbf{94.00}  & \multicolumn{1}{c|}{}                         & \multicolumn{1}{l|}{2nd}                 & \multicolumn{1}{c|}{76.23}  & \multicolumn{1}{c|}{\textbf{86.42}}  & \multicolumn{1}{c|}{77.67}  \\ \cline{2-5} \cline{7-10}
                         & 3rd                  & 84.10   & 87.21  & \textbf{100.00} & \multicolumn{1}{c|}{}                         & \multicolumn{1}{l|}{3rd}                  & \multicolumn{1}{c|}{\textbf{91.69}}  & \multicolumn{1}{c|}{82.15}  & \multicolumn{1}{c|}{89.31}  \\ \hline
\multirow{3}{*}{\#15}    & 1st                  & \textbf{100.00}  & 68.79  & 67.34  & \multicolumn{5}{c}{\multirow{3}{*}{}}                                                                                                                                                 \\ \cline{2-5}
                         & 2nd                 & 85.12   & \textbf{91.26}  & 75.14  & \multicolumn{5}{c}{}                                                                                                                                                                  \\ \cline{2-5}
                         & 3rd                  & 66.47   & 70.16  & \textbf{80.35}  & \multicolumn{5}{c}{}                                                                                                                                                                  \\ \cline{1-5}
\end{tabular}
\end{center}
\footnotesize{'1st', '2nd', and '3rd' mean the data obtained from the first, second, and third experiments of one participant, respectively.}
\end{table*}

\begin{table}[!h]\normalsize
\caption{The average accuracies (\%) of our emotion model across sessions.}\label{tab:table9}
\begin{center}
\begin{tabular}{|c|l|c|c|c|}
\hline
\hline
\multirow{2}{*}{Stats.} & \multirow{2}{*}{Train} & \multicolumn{3}{c|}{Test} \\ \cline{3-5}
                       &                        & First  & Second  & Third  \\ \hline
\multirow{3}{*}{Mean}  & First                  & \textbf{90.83}  & 72.55   & 67.22  \\ \cline{2-5}
                       & Second                 & 75.86  & \textbf{88.22}   & 76.62  \\ \cline{2-5}
                       & Third                  & 76.28  & 78.17   & \textbf{87.80}  \\ \hline
\multirow{3}{*}{Std.}  & First                  & \textbf{8.64}   & 10.29   & 10.42  \\ \cline{2-5}
                       & Second                 & \textbf{7.71}   & 8.59    & 15.34  \\ \cline{2-5}
                       & Third                  & 11.47  & 13.41   & \textbf{10.97}  \\ \hline\hline
\end{tabular}
\end{center}
\end{table}

The results are presented in Tables \ref{tab:table8} and \ref{tab:table9}. From the mean accuracies and standard deviations, we find that the accuracies with training set and test set from the same sessions are much higher than those from different sessions. The performance of the emotion recognition model is better with train data and test data obtained from sessions performed in nearer time. A comparative mean classification accuracy of 79.28\% is achieved using our emotion recognition model with training and testing datasets from different sessions. This result implies that the relation between the variation of emotional states and the EEG signal is stable for one person in a period of time. With the passage of time, the performance of the model may become worse. So the adaption of the computational model should be further studied as future work.

Now we consider the situation of cross-subject and examine the subject-independent emotion recognition model. Here, we employ a leave-one-out cross validation to investigate the classification performance in a subject-independent approach and use linear SVM classifier with DE features from five frequency bands as inputs. The average accuracy and standard deviation with subject-independent features reach 60.93\% and 13.95\%, respectively. These results indicate that the subject-independent features are relatively stable and it is possible to build a common emotion recognition model. But on the other hand, the factors of individual difference should be considered to build a more robust affective computing model.

We have investigated how stable our emotion recognition model both across subjects and sessions, and we find that the performance of the model across subjects and sessions are worse than that on single experiment. In general, we want to train the model on the EEG data from a set of subjects or sessions and make inference on the new data from other unseen subjects or sessions. However, it is technically difficult due to individual differences across subjects with the inherent variability of the EEG measurements such as environmental variables \cite{olivetti2014meg}. Although different emotions share some commonalities of neural patterns as we report above, they still contains some individual differences for different subjects and different sessions, which may lead to the changes of underlying probability distribution from subject to subject or from session to session. This is why the average accuracy of the classifiers trained and tested on each individual subject or session is much higher than that of a classifier trained on a set of subjects or sessions and tested on other subjects or sessions.

\section{Conclusions and Future Work}
In this paper, we have systematically evaluated the performance of different popular feature extraction, feature selection, feature smoothing and pattern classification methods for emotion recognition on both our SEED dataset and the public DEAP dataset. From the experimental results, we have found that GELM with the differential entropy features outperforms other methods. For DEAP dataset, the best average classification accuracy of 69.67 percent for quadrants of VA space with 32 participants is obtained using 5-fold cross-validations. For our SEED dataset, the best average classification accuracy of 91.07 percent from 45 experiments is obtained using 5-fold cross-validations. The comparative classification accuracies achieved show the reliability and superior performance of our machine learning methods in comparison with the existing approaches. We have utilized these methods to investigate the stability of neural patterns over time.


On our SEED dataset, an average classification accuracy of 79.28\% is achieved with training and testing datasets from different sessions. The experimental results indicate that neural signatures and stable EEG patterns associated with positive, neutral and negative emotions do exist. We have found that the lateral temporal areas activate more for positive emotion than negative one in beta and gamma bands; the neural patterns of neutral emotion have higher alpha responses at parietal and occipital sites; and the negative emotion patterns have significant higher delta responses at parietal and occipital sites and higher gamma responses at prefrontal sites. The experiment results also indicate that the stable EEG patterns across sessions exhibit consistency among repeated EEG measurements of the same subject.

In this study, we investigate the stable neural patterns of three emotions, positive, neutral and negative. For the future work, more categories of emotions will be studied and the generalization of our proposed approach extended to more categories of emotions will be evaluated. Moreover, different factors such as gender, age, and race should be considered. To make automatical emotion recognition models adaptable, the factors like individual difference and temporal evolution should be considered. One possible way to dealing with these problems is to adopt transfer learning techniques \cite{pan2010survey,samek2013transferring,zheng2015transfer}.

\ifCLASSOPTIONcompsoc
  \section*{Acknowledgments}
\else
  \section*{Acknowledgment}
\fi

The authors wish to thank all the participants in the emotion experiments and thank the Center for Brain-Like Computing and Machine Intelligence for providing the platform for EEG experiments. This work was supported in part by the grants from the National Natural Science Foundation of China (Grant No. 61272248), the National Basic Research Program of China (Grant No. 2013CB329401), and the Science and Technology Commission of Shanghai Municipality (Grant No.13511500200).

\ifCLASSOPTIONcaptionsoff
  \newpage
\fi



\bibliographystyle{IEEEtran}
\bibliography{bare_jrnl_compsoc}

\end{document}